# Modification of emission properties of ZnO layers due to plasmonic near-field coupling to Ag nanoislands

Joanna Papierska[1], Bartłomiej S. Witkowski[3], Anastasiya Derkachova[3], Krzysztof P. Korona[1], Johannes Binder[1], Krzysztof Gałkowski[1], Łukasz Wachnicki[3], Marek Godlewski[3], Tomasz Dietl[2,3] and Jan Suffczyński[1]

[1] *Institute of Experimental Physics, Faculty of Physics, University of Warsaw, Hoża 69 St., PL-00-681 Warszawa, Poland*

[2] *Institute of Theoretical Physics, Faculty of Physics, University of Warsaw, Hoża 69 St., PL-00-681 Warszawa, Poland*

[3] *Institute of Physics, Polish Academy of Sciences, Lotników 32/46 Av., PL-02-668 Warszawa, Poland*

E-mail: Jan.Suffczynski@fuw.edu.pl

**Abstract** A simple fabrication method of Ag nanoislands on ZnO films is presented. Continuous wave and time-resolved photoluminescence and transmission are employed to investigate modifications of visible and UV emissions of ZnO brought about by coupling to localized surface plasmons residing on Ag nanoislands. The size of the nanoislands, determining their absorption and scattering efficiencies, is found to be an important factor governing plasmonic modification of optical response of ZnO films. The presence of the Ag nanoislands of appropriate dimensions causes a strong (threefold) increase in emission intensity and up to 1.5 times faster recombination. The experimental results are successfully described by model calculations within the Mie theory.

***Keywords*** zinc oxide, silver nanoislands, photoluminescence, Mie theory, exciton

## Introduction

Recent intensive research in the domain of plasmonics has shown that optical effects in semiconductors are strongly modified by coupling of light to plasma oscillations in metallic nanostructures residing on or in the proximity of the semiconductor surface [1, 2]. In particular, a considerable enhancement of the emission intensity in the case of either self-organized [3] or lithographically defined metal nanostructures [1] was observed. Many factors such as density, shape, size and its distribution, agglomeration, internal structure, as well as composition and porosity of the surface of metal nanostructures determine their impact on the optical response of semiconductors [4,1]. A number of publications on ZnO-metal hybrid structures devoted to different aspects of the relevant physics have appeared within the last decade. In particular, continuous wave photoluminescence (PL) studies of Ag/ZnO demonstrated an enhancement of subband gap emission by Ag nonislands [4,5]. It was speculated that this enhancement is due the resonant coupling of spontaneous emissions in ZnO to the surface plasmons in metal overlayers [4,5].

Here, we apply the known [4] and an innovative method for fabrication on ZnO thin films of silver (Ag) islands with diameter of 10-60 nm and 60-120 nm, respectively. By employing photoluminescence (PL), time resolved PL and transmission measurements we determine experimentally the impact of the Ag coating on the ZnO films optical response in the visible and ultra-violet range. We describe the set of our experimental data within the Mie theory. The employed model shows that the size and resulting efficiencies of the light scattering and absorption by metal nanoparticles is decisive for modifying optical performance of semiconductors.

## Samples

The studied ZnO layers are grown by atomic layer deposition (ALD) on a 3 μm GaN template [6]. The ZnO layers of 50 nm thickness are studied in order to assure that the whole layer experiences the plasmonic mode decaying within the range of tens of nanometers from the sample surface [7].

Two methods have been employed to cover samples by Ag nanoislands. Sample A is covered with silver islands by sputter deposition lasting for 2 s. Sample B is prepared in a more elaborate way, exploring an original method of metal island formation. The film surface is first coated by Ag sputtering over 15-20 s, yielding a 3 nm layer of Ag, and it is then annealed at a temperature of 750°C for 3 min. The annealing process results in the formation of metal islands. Actually, in the case of both samples half of the film is masked in a way preventing Ag deposition. This part of the ZnO surface, uncovered by Ag nanoislands, serves as a reference. As determined by scanning electron microscope (SEM) measurements (Fig. 1), islands of diameter distributed between D = 10 - 60 nm (with dominating D = 10 nm) with a surface density of $\rho \approx$ 2000/μm$^2$ are obtained by the first method (sample A). In the case of the second method islands of diameter distributed equally between 60 nm and 120 nm with a surface dentisity of $\rho \approx 25/\mu m^2$ are obtained (sample B). The Ag islands have a quasi-spherical shape in both samples.

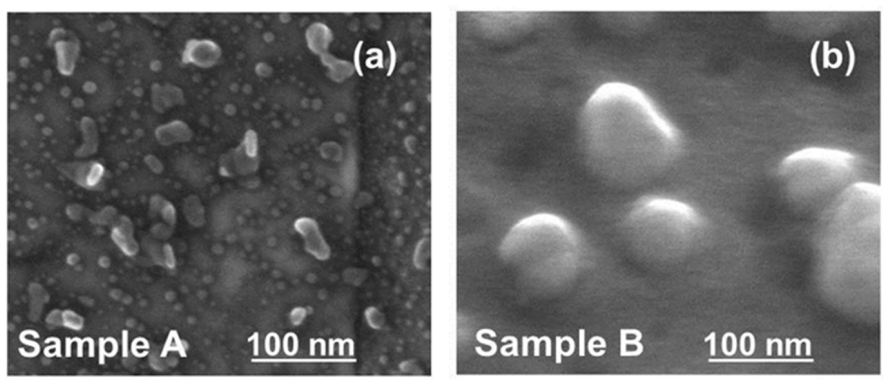

Fig. 1 SEM image of sample A (a) and sample B (b). The magnitudes of diameter and surface density of Ag nanoislands are determined to be 10-60 nm and 2000/$\mu m^2$ (sample A) or 60-120 nm and 25/$\mu m^2$ (sample B), respectively.

The presence of Ag nanoislands on the surfaces of both samples is confirmed by Energy-dispersive X-ray (EDX) measurements conducted with a beam of energy of 5 keV. A comparison of the EDX profiles (see Fig. 2) measured on the representative points 1 and 2 on the sample B (see inset to Fig. 2) shows that a clear maximum at around 3 eV related to Ag is evidenced only in the case of a sample region covered with the nanoisland. The Ag maximum is assisted by a weak satellite maximum at around 2.3 eV related to sulfide, what is presumably due to a passivation of nanoislands surface by sulfite atoms from the air. Due to the strong background signal related to oxygen from ZnO it is impossible to determine the degree of oxidation of the Ag nanoislands.

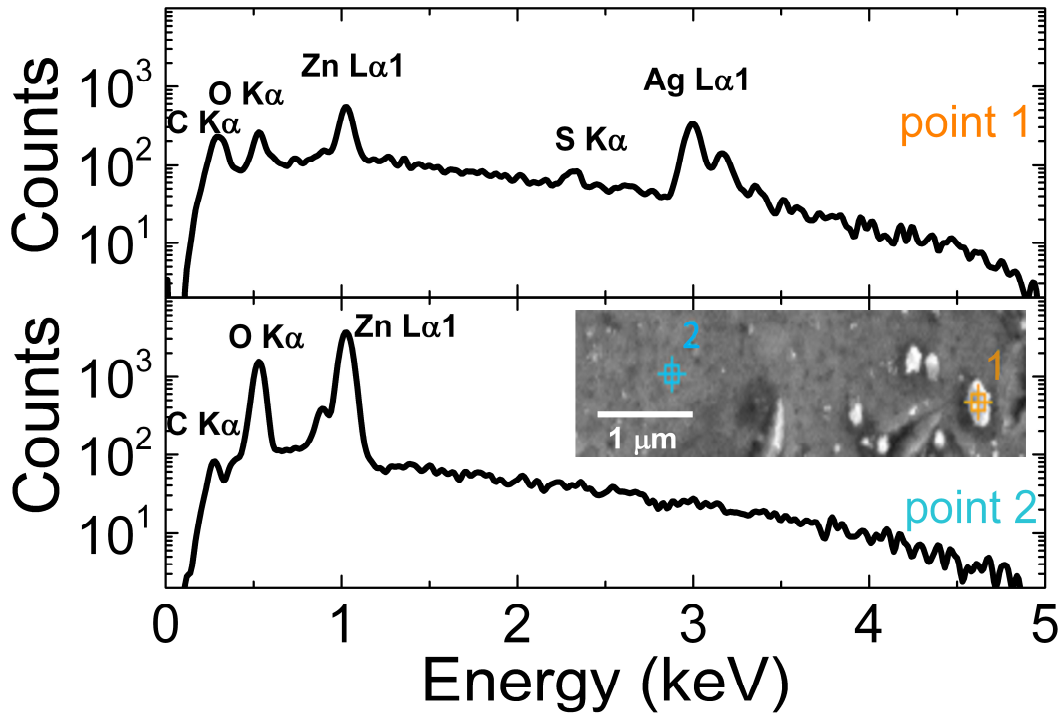

Fig. 2 EDX spectra measured at two points on sample B: with (point 1) and without (point 2) Ag nanoisland. The inset: SEM image of the sample region with indicated points 1 and 2.

## Experiment

Photoluminescence, time-resolved PL, and transmission spectra are measured at temperatures ranging from 4 to 300 K. The emission is excited at 325 nm (He-Cd laser) or at 300 nm (frequency tripled Ti:Sapp pulsed laser) in the case of PL and time resolved PL, respectively. A halogen lamp serves as the light source in the transmission experiment. The incident beam is focused on the sample surface to a spot of 0.5 mm (in the case of time resolved PL 3 µm) diameter. The signal arising from the sample is dispersed by a grating monochromator (600 gr/mm or 1800 gr/mm) and recorded by a CCD camera or a streak camera in the case of time resolved PL.

## Results

In order to characterize optical properties of the hybrid structures, the transmission in the energy range 1.5 – 3.6 eV has been measured. We estimate that for the thickness and the absorption coefficient [8] of the ZnO layer, its absorption is negligible. Accordingly, for energies below the band gap of the GaN template (~3.42 eV at 300 K), the observed extinction results from the interaction of light with the Ag nanoislands. The ratio of the transmission of the sample B part covered with Ag nanoislands to the transmission of the uncovered part exhibits a wide (0.6 eV) minimum (modulation 60%) at around ~2.3 eV, indicating the position of the extinction maximum. In the case of sample A, the extinction related to the Ag nanoislands is much weaker and shows no clear minimum.

Continuous-wave PL spectra taken for both studied samples at 5 K are presented in Figs. 3(a) and 3(c). The spectra collected for Ag covered and uncovered regions are shown by solid and dotted lines, respectively. The dominating transition at ~3.36 eV is attributed to bound exciton emission in ZnO. The signal observed in the ~2.90 - 3.33 eV energy range is assigned to stacking faults. A weak defect related signal appears below ~2.9 eV. A more detailed analysis of the near-band-gap emission is presented below. As expected and seen in Figs. 3(a) and 3(c), excitonic emission shows a shift towards lower energies and broadening when temperature is increased to 300 K.

In order to determine the impact of Ag nanoparticles on the emission from ZnO films, the ratio of the photoluminescence from the covered to uncovered part of the sample is evaluated for both samples (Figs. 3(b) and 3(d)). As seen in Fig. 3(b), the presence of 10-60 nm Ag islands affects the PL signal of sample A only weakly, leading to its slight decrease in the almost entire registered spectral range. Much more pronounced effects are observed in the case of sample B, covered with 60-120 nm Ag islands. As seen in Fig. 3(d), the presence of Ag nanoparticles exhibits up to threefold emission enhancement in the spectral range below ~2.4 eV. However, in the spectral range above ~2.4 eV including the excitonic region, a quenching of the emission is observed. A comparison of the results for 2 K and 300 K shown in Fig. 3 indicates that the observed effects of the PL enhancement or quenching are basically temperature independent. This is understood, provided that their origin is related to plasmonic effects in Ag, which are governed by the magnitude of electron density, independent of temperature. A model presented in "Theoretical model" section corroborates the plasmonic origin of the above observations.

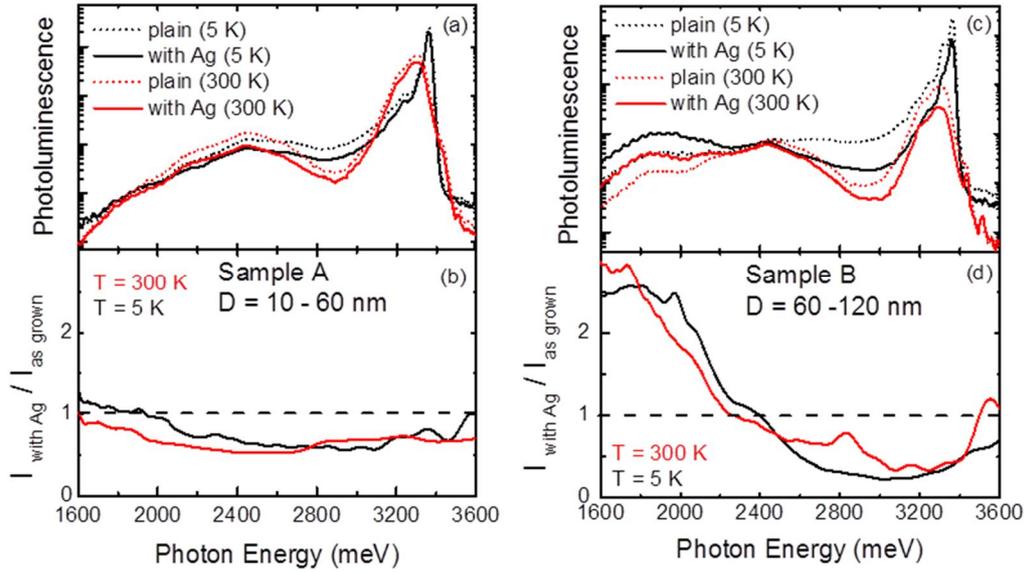

Fig. 3 Photoluminescence spectra of the ZnO layers: sample A (a) and sample B (c) as grown (solid line) and covered with Ag nanoparticles (dashed line). Ratio of the photoluminescence intensity from the covered to uncovered parts of sample A (b) and B (d). The spectra registered at T = 5 K and 300 K are indicated in black and red (or grey), respectively.

Results of time-resolved PL measurements at 4 K shed more light on the nature of the signal quenching in the near-band-gap spectral region resulting from the presence of the Ag nanoparticles. In Fig. 4, a temporal evolution of the PL spectra following the excitation pulse is shown for both covered and uncovered parts of sample B. A few dominant peaks are present in the signal coming from both parts of the sample. As indicated in Fig. 4, free exciton (FX) and donor bound exciton (DX) transitions are observed at 3.375 eV and 3.361 eV, respectively [9]. Two weak peaks are observed at 3.33 eV and 3.34 eV; their origin is not clear. Stacking fault (SF) related emission is observed in the spectral range 3.30 – 3.33 eV [10]. Furthermore, acceptor bound exciton (AX) [9], not observed in the signal from the uncovered sample, appears at 3.356 eV in the emission from the Ag covered sample. This results most probably from the diffusion of the Ag atoms, acting as acceptor centers, to the ZnO layer during the process of the sample annealing.

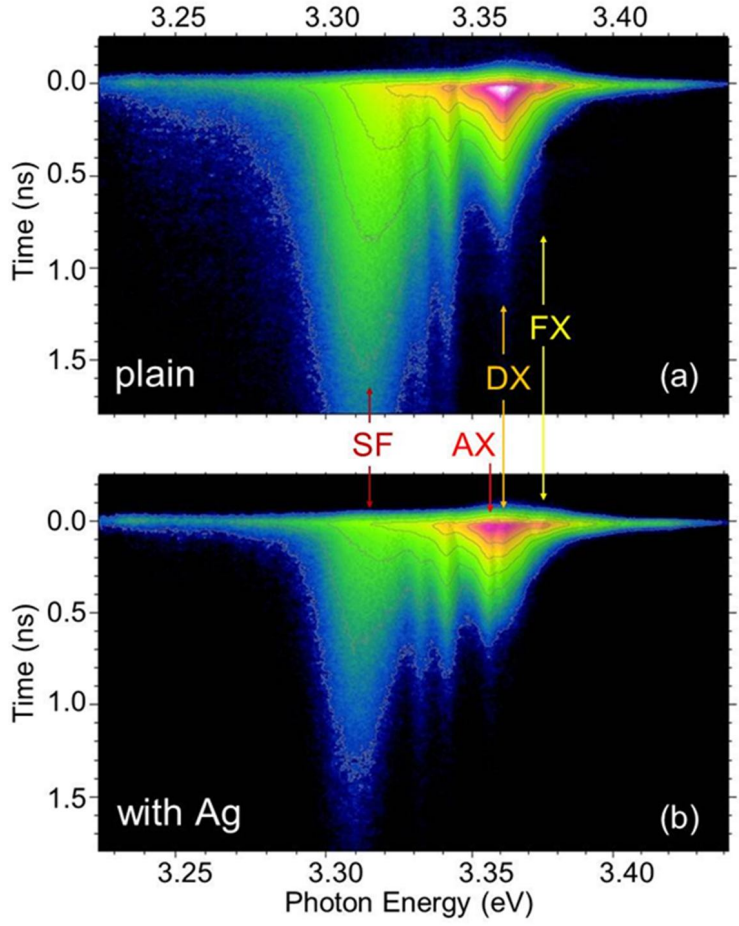

Fig. 4 Time resolved PL spectra taken for sample B at 4 K for regions: plain (a) and covered (b) by Ag nanoislands of diameter D = 60 - 120 nm. The contour values are set by an exponential rule, and the ratio between the two nearest contour curves is r = $10^{1/2}$.

The analysis of time–resolved PL for sample B (see Fig. 4) reveals that the decay of free excitons is monoexponential and relatively fast (25 ± 5 ps and 35 ± 5 ps, for the Ag-covered and uncovered region, respectively). The DX decays are composed of two components. The shorter decay constant $\tau_1$ is 66 ± 5 ps and 77 ± 5 ps for the Ag-covered and uncovered regions, respectively. The longer component has the lifetime $\tau_2$ of about 0.64 ± 0.05 ns on both sides of the sample. Also the longer components of the decay of SF, as well as the peaks at 3.33 eV and the 3.34 eV have similar dynamics on both sides of the sample and their lifetimes are 0.61 ± 0.05 ns, 0.69 ± 0.05 ns and 0.54 ± 0.05 ns, respectively. The AX also has two exponential decays with $\tau_1$ = 80 ± 5 ps and $\tau_2$ = 0.95 ± 0.05 ns.

The values of the FX decay constants and the shorter component of the decay $\tau_1$ for the DX and the SF are collected in table 1. Intensity values of time integrated PL, averaged over a few points on the sample, are also displayed. A comparison of the values presented in table 1 obtained for covered and uncovered parts of the sample indicates that the FX-related decay times are reduced in the presence of the Ag nanoislands. Also the emission intensity is lower for the covered side. These effects are attributed to the existence of an additional nonradiative channel related to the coupling of the emission to localized surface plasmons confined in the Ag nanoislands. A similar reduction of the decay time was previously observed [2], however in that case an additional channel was of a radiative nature and led to an increase of the emission magnitude. As modeling presented in Sec. V reveals, the damping of the emission magnitude observed in our case results from the fact that for Ag nanoparticles with diameters of 80 - 90 nm (or smaller) the absorption efficiency dominates over the scattering efficiency in the near-band-gap spectral range. The Ag nanoislands influence mainly free excitons, either because they can diffuse to the surface or because spatially extended excitations are more strongly coupled to plasmons. Accordingly, the recombination rate of bound excitons is only weakly affected by Ag coating. However the PL intensity of bound excitons is reduced, as Ag nanoislands, by constituting centers of free exciton recombination, diminish the concentration of bound excitons. Time-resolved measurements on sample A show that appropriately small Ag nanoislands have a minor effect on the PL lifetimes.

Table 1 PL lifetimes (fast components) and intensities determined for free exciton (FX), bound exciton (DX), and stacking fault (SF) transitions in sample B.

| Transition | Energy (eV) | PL lifetime (ps) | | Intensity ($10^3$ counts) | |
| --- | --- | --- | --- | --- | --- |
| | | plain | with Ag | plain | with Ag |
| FX | 3.375 | 35 ± 5 | 25 ± 5 | 4.7 | 4.0 |
| DX | 3.36 | 77 ± 5 | 66 ± 5 | 39 | 17 |
| SF | 3.30 – 3.33 | 40 ± 15 | 30 ± 10 | 7.0 | 1.3 |

## Theoretical model

In order to describe the effects observed experimentally we employ a model developed within the Mie scattering theory [11-13]. According to this theory, the efficiencies $Q_{scat}$, $Q_{ext}$, $Q_{abs}$ of respectively scattering, extinction and absorption of the light interacting with a single spherical particle of a radius R, are (notation according to [13]):

$$Q_{scat} = \frac{\lambda^2}{2\pi^2 R^2}\sum_{n=1}^{\infty}(2n+1)(|a_n|^2 + |b_n|^2), \qquad (1)$$

$$Q_{ext} = \frac{\lambda^2}{2\pi^2 R^2}\sum_{n=1}^{\infty}(2n+1)Re\{a_n + b_n\}, \qquad (2)$$

$$Q_{abs} = Q_{ext} - Q_{scat}, \qquad (3)$$

with:

$$a_n = \frac{m\psi_n(mx)\psi_n'(x) - \psi_n(x)\psi_n'(mx)}{m\psi_n(mx)\xi_n'(x) - \psi_n(x)\psi_n'(mx)}, \qquad (4)$$

$$b_n = \frac{\psi_n(mx)\psi_n'(x) - m\psi_n(x)\psi_n'(mx)}{\psi_n(mx)\xi_n'(x) - m\xi_n(x)\psi_n'(mx)}, \qquad (5)$$

where

$$x = k_{out}R = \frac{2\pi n_{out}R}{\lambda},\ m = \frac{k_{in}}{k_{out}} = \frac{n_{in}}{n_{out}} = \frac{\sqrt{\varepsilon_{in}}}{\sqrt{\varepsilon_{out}}},\ k_{in} = \frac{\omega}{c}\sqrt{\varepsilon_{in}},\ k_{out} = \frac{\omega}{c}\sqrt{\varepsilon_{out}},$$

with $\varepsilon_{in}$ and $\varepsilon_{out}$ being the dielectric functions of the particle and medium respectively. The prime indicates differentiation with respect to a function argument. $\psi_n(z)$ and $\xi_n(z)$ are Riccati-Bessel spherical functions which can be expressed by the Bessel $J_{n+\frac{1}{2}}(z)$, Hankel $H^{(1)}_{n+\frac{1}{2}}(z)$ and Neuman $N_{n+\frac{1}{2}}(z)$ cylindrical functions of the half order.

The dielectric function of silver is assumed as in the classical Drude model taking into account the surface scattering and interband transitions:

$$\varepsilon_{in} = \varepsilon_{ib} - \frac{\omega_p^2}{\omega^2 - i\gamma\omega}, \qquad (6)$$

$$\gamma = \gamma_{bulk} + A\frac{v_F}{R}, \qquad (7)$$

where $\omega_p$ is the bulk plasmon frequency, $\gamma_{bulk}$ is a phenomenological relaxation constant of bulk silver, $v_F$ is the Fermi velocity, A is a dimensionless parameter and $\varepsilon_{ib}$ is the phenomenological parameter typically smaller than unity [14] describing a contribution of bound electrons to polarizability [15].

The Drude dielectric function parameters, $\varepsilon_{ib}$, $\gamma_{bulk}$ and $\omega_p$ play an important role in theoretical predictions of the light/metal nanoparticles interaction. They define the position and width of the peak in extinction, scattering and absorption efficiencies associated with the surface plasmon resonance in metal nanoparticles. Usually, the effective parameters of Drude dielectric function of pure silver, $\varepsilon_{ib} = 3.7$, $\gamma_{bulk} = $ 0.18 eV and $\omega_p$ = 9.1 eV, provide a good description of the experimental data [16] and we use them in the case of calculations for sample A. However, annealing of sample B at temperature of 750°C leads presumably to interdiffusion of Ag and Zn. As a result, the effective parameters might be modified, as already noted in several other cases [17-21]. In the present work, in order to describe our results for sample B we use $\varepsilon_{ib}$ = 8.1, $\gamma_{bulk}$ = 0.3 eV and $\omega_p$ = 9.1 eV. These parameters assure a good agreement of the calculated extinction position and width with the one determined from the transmission experiment on sample B (see Figs. 5(a) and 5(b)).

The effective complex refractive index of the nanoparticle surrounding $n_{out}$ is defined by the "mixture rule" [22] as:

$$n_{out} = fn_{Air} + (1-f)n_{ZnO}, \tag{8}$$

where f = 0.66 is a parameter resulting from a quasi-spherical shape of nanoislands. Furthermore, we take $n_{Air}$ = 1 and $n_{ZnO} = \sqrt{\varepsilon_{ZnO}}$, where the ZnO dielectric function $\varepsilon_{ZnO}$ is calculated following Ref. [8].

In Fig. 5 the calculated spectral distributions of extinction, scattering and absorption are shown. The calculation is made assuming that Ag nanoislands in the case of sample A have diameters D distributed from 10 nm to 60 nm every 10 nm, where D = 10 nm is taken with 0.5 weight and other diameters with equal 0.1 weight. In the case of sample B, diameter distribution from 60 nm to 120 nm every 10 nm with equal weight is assumed. As expected, the extinction maximum shifts to lower energies when the radius of the sphere increases: the computed energy of extinction maximum is at ~3.2 eV (sample A, Fig. 5(a)) and at ~2.9 eV or ~2.4 eV (sample B, Figs. 5(b) and 5(c)) depending whether the dielectric function of pure Ag or the modified one is assumed, respectively. A comparison of the dashed curves in Fig.5 affirms that an increase of the sphere radius results also in a decrease of the absorption efficiency and in an enhancement of the scattering efficiency.

In order to link these results to experimental findings, we recall that plasma oscillations excited by the incident light decay by nonradiative and radiative processes are described by absorption and scattering efficiencies, respectively. With this in mind, the computed relative intensities of absorption and scattering efficiencies explain PL and time resolved PL experiments results. In particular, as seen in Fig.5(a), the absorption dominates over the scattering in the whole studied spectral range in the case of sample A, leading to a decrease of the PL intensity, as observed (Figs. 3(a) and 3(b)).

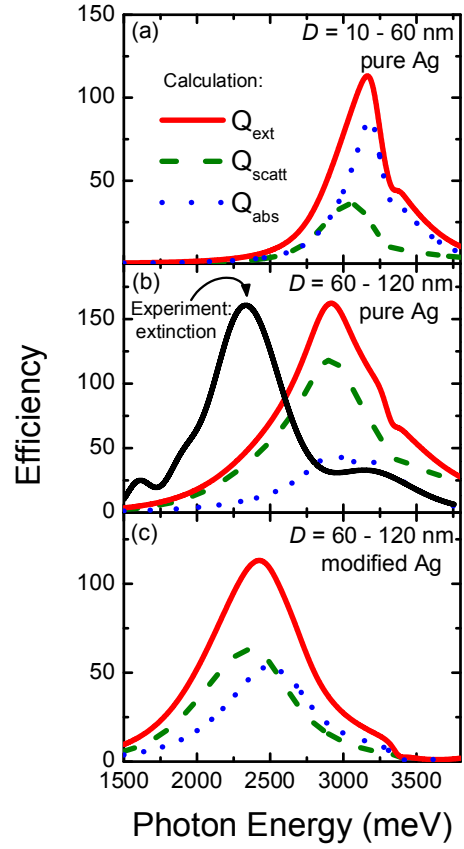

Fig. 5 Efficiency of extinction (Qext), scattering (Qscat) and absorption (Qabs) calculated using Mie theory for silver nanoparticles with a diameter of D = 10 - 60 nm (a) and D = 60 - 120 nm (b-c) deposited on ZnO at 300 K. Pure silver dielectric function parameters: $\varepsilon_{ib}$ = 3.7, $\gamma_{bulk}$ = 0.18 eV and $\omega_p$ = 9.1 eV were assumed in (a) and (b), while modified ones: $\varepsilon_{ib}$ = 8.1, $\gamma_{bulk}$ = 0.3 eV and $\omega_p$ = 9.1 eV in (c). The extinction determined from the transmission experiment on sample B (D = 60 - 120 nm) and normalized to the maximum value of the respective calculated extinction is shown in panel (b).

In contrast to sample A, in the case of sample B the absorption efficiency dominates over the scattering efficiency only for energies above 2.5 eV, see Fig. 5(c). As a result, a quenching of the emission is observed for energies above ~2.5 eV (see Figs. 3(c) and 3(d)), where excitonic transitions are visible (see Fig. 4 and Table 1). However, below ~2.5 eV a substantial emission enhancement occurs (see Figs. 3(c) and 3(d)).

## Conclusions

In this paper, the plasmonic coupling in hybrid system ZnO thin film Ag nonoislands has been investigated by continuous wave and time-resolved spectroscopy. The fabrication process, involving sputtering and annealing, has served to obtain Ag nanoislands of dimensions insuring the presence of strong plasmonic effects. The temperature independent enhancement or quenching of the emission is observed and explained by the numerical model developed within the Mie theory. The ratio of the absorption to scattering efficiencies is found to govern the emission enhancement or quenching. The model highlights the impact of the sample preparation on the dielectric function of Ag. The results may be useful for applications requiring a modification of the optical response of a semiconductor in the UV and visible spectral region. In particular, many groups have reported recently on the fabrication, by crystallographic or chemical phase separation, of hybrid structures consisting of magnetic metallic nanocrystals embedded in a semiconductor matrix [23]. Our results indicate dimensions of metallic nanocrystals that could lead to either enhancement or quenching of emission intensity in such heterogeneous systems.

**Acknowledgments** The work was supported by the European Research Council through the FunDMS Advanced Grant (#227690) within the ''Ideas'' 7th Framework Programme of the EC, NCBiR project LIDER, and InTechFun (Grant No. POIG.01.03.01-00-159/08). We thank P. Kossacki for critical reading of the manuscript.